\documentclass[aps,prl,twocolumn,superscriptaddress,showpacs]{revtex4}
\usepackage{graphicx}
\begin{document}
\title{Symmetry of Multiferroicity in a Frustrated Magnet TbMn$_2$O$_5$}
\author{J. Okamoto}
\affiliation{National Synchrotron Radiation Research Center,
Hsinchu 30076, Taiwan}
\author{D. J. Huang}
\altaffiliation[E-mail address:] {\emph{ djhuang@nsrrc.org.tw}}
\affiliation{National Synchrotron Radiation Research Center,
Hsinchu 30076, Taiwan} \affiliation{Department of Physics,
National Tsing Hua University, Hsinchu 30013,
Taiwan}\affiliation{Department of Electrophysics, National
Chiao-Tung University, Hsinchu 30010, Taiwan}
\author{C.-Y. Mou}
\affiliation{Department of Physics, National Tsing Hua University,
Hsinchu 30013, Taiwan}
\author{K. S. Chao}
\affiliation{Department of Electrophysics, National Chiao-Tung
University, Hsinchu 30010, Taiwan}
\author{H.-J. Lin}
\affiliation{National Synchrotron Radiation Research Center,
Hsinchu 30076, Taiwan}
\author{S. Park}
\author{S-W. Cheong}
\affiliation{Rutgers Center for Emergent Materials and Department
of Physics \& Astronomy, Rutgers University, Pitcataway, New
Jersey 08854, USA}

\author{C. T. Chen}
\affiliation{National Synchrotron Radiation Research Center,
Hsinchu 30076, Taiwan}
\date{\today}
\begin{abstract}
Based on measurements of soft x-ray magnetic scattering and
symmetry considerations, we demonstrate that the magnetoelectric
effect in TbMn$_2$O$_5$ arises from an internal field determined
by $\vec{S}_{\vec{q}} \times {\vec{S}_{-\vec{q}}}$ with
$\vec{S}_{\vec{q}}$ being the magnetization at modulation vector
$\vec{q}$, whereas the magneto-elastic effect in the exchange
energy governs the response to external electric fields. Our
results set fundamental symmetry constraints on the microscopic
mechanism of multiferroicity in frustrated magnets.
\end{abstract}
 \maketitle

Materials which exhibit coexistence of magnetism and
ferroelectricity with cross coupling, termed multiferroicity, are
attractive because they offer the possibility for realizing mutual
control of electric and magnetic properties. The key phenomenon
behind such mutual control lies on the capability for the
induction of magnetization by an electric field or of electric
polarization by a magnetic field, known as the magnetoelectric
(ME) effect \cite{Fiebig,Eerenstein}. The ME effect is an
important characterization of multiferroicity but has been poorly
understood. The effect could be largely enhanced by the presence
of internal fields. However, such enhancement requires the
coexistence and strong coupling of magnetism and ferroelectricity
(FE), which rarely happen in real materials. Recent discoveries of
giant magnetoelectric couplings in frustrated magnets
\cite{Kimura,Hur} thus offer new opportunities for a thorough
scientific understanding of multiferroicity as well as
multiferroic applications.

In frustrated magnets, such as $R$MnO$_3$ and $R$Mn$_2$O$_5$ ($R$
= Tb, Dy, and Ho)
\cite{Kimura,Hur,Hur-2,Higashiyama,Kobayashi,Chapon,Chapon_1,Blake,Higashiyama05,Kenzelmann,Arima},
the spontaneous electric polarization ($\vec{P}$) appears in
certain antiferromagnetic (AF) phases. Unlike old examples of
multiferroics, the magnetoelectric couplings exhibited by these
materials are gigantic, and the magnetic phases involved are
complicated and commonly incommensurate with lattice. The magnetic
transition temperature is higher than the ferroelectric one,
suggesting that the ferroelectricity is induced by magnetic order.
Furthermore, the inversion symmetry in the magnetic phases with
ferroelectricity is broken \cite{Kenzelmann,Arima}, implying that
the magnetic order couples to odd orders of $\vec{P}$. In
addition, these magnets show anomalies in the temperature
dependence of dielectric constant $\varepsilon$. For $R$MnO$_3$,
the ferroelectric transition is accompanied by a magnetic
transition from incommensurate sinusoidal to spiral AF order
\cite{Kenzelmann,Arima}. Kenzelmann \emph{et al}. have applied the
Ginzburg-Landau theory to understand the multiferroic behavior
\cite{Kenzelmann}. In contrast, although Chapon \emph{et al}.
\cite{Chapon_1} found that the ferroelectricity in YMn$_2$O$_5$
results from acentric spin-density waves, little is known about
the underlying mechanism of multiferroicity in $R$Mn$_2$O$_5$
because of their structural complexity. The exact relation and
interplay between AF order and ferroelectricity in frustrated
magnets are unknown and remain controversial
\cite{Arima,Efremov,Chapon_1,Katsura,Sergienko}.

The presence of internal fields is the simplest origin of the
cross coupling between magnetism and ferroelectricity.
Microscopically, however, it is difficult to identify them due to
their weak effect. Historically, scattering has been shown to be
powerful for measuring accumulated microscopic changes and for
revealing orderings and their relations directly. Neutron
scattering \cite{Shull}, for example, first convincingly proved
the existence of the antiferromagnetic phase of MnO. Here, by
resorting to soft x-ray magnetic scattering \cite{Hannon} and an
analysis based on the phenomenological Ginzburg-Landau approach,
we demonstrate that the magnetically induced ferroelectricity in
TbMn$_2$O$_5$ arises from an internal field determined by
$\vec{S}_{\vec{q}} \times \vec{S}_{-\vec{q}}$ with
$\vec{S}_{\vec{q}}$ being the magnetization at modulation vector
$\vec{q}$. The results indicate that the non-collinearity of spins
is essential for the existence of induced $\vec{P}$, independent
of commensurability and chirality. In contrast, the external
electric fields alter the exchange coupling, yielding anomalies in
the temperature dependence of the dielectric constant.

\begin{figure}
\includegraphics[width=8.5cm]{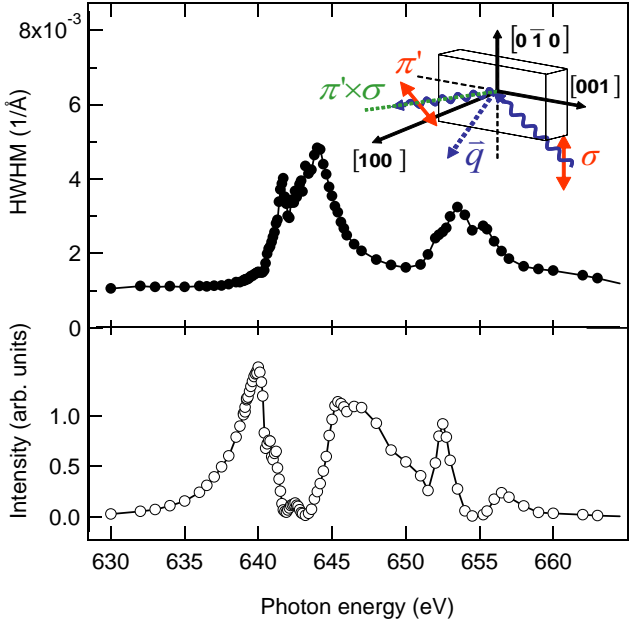}
\caption{(Color) Photon-energy dependence of HWHM and scattering
intensity of TbMn$_2$O$_5$ with $\vec{q}$ =
(\(\frac12\),0,\(\frac14\)) and $\sigma$ polarization at 30 K. The
intensity was obtained by fitting the $q$ scan at each photon
energy with a Lorentzian function and a linear background. Inset:
Scattering geometry with $\sigma$ and $\pi'$ polarizations for
incident and scattered photons, respectively. With $\vec{q}$ near
$(\frac{1}{2}, 0, \frac{1}{4})$, the angle between the scattered
x-ray and the $a$ axis is $\sim 6^\circ$ and that between the
incident x-ray and the $c$ axis is $\sim 19^\circ$.} \label{Fig1}
\end{figure}

We present measurements of soft x-ray magnetic scattering around
the $L_3$ ($2p_{3/2}\rightarrow3d$) absorption edge of Mn to
reveal the detailed coupling of ferroelectricity and AF order in
TbMn$_2$O$_5$. Soft x-ray magnetic scattering is a newly developed
technique which is sensitive to the magnetic moment of
transition-metal $d$ electrons
\cite{Hannon,Wilkins,Thomas,Stojic}, allowing us to probe magnetic
order with high sensitivity. The scattering amplitude is
proportional to the magnetization $\vec{S}_{\vec{q}}$, which is $
\sum_j \vec{S}_j {\rm e}^{i \vec{q} \cdot \vec{r_j}}$ with
$\vec{S}_j$ and $\vec{r}_j$ being spin moments and position
vectors, respectively. We performed scattering measurements with
the elliptically polarized-undulator beamline at National
Synchrotron Radiation Research Center (NSRRC) in Taiwan.  The
inset of Fig. 1 illustrates a schematic view of scattering
geometry. A single crystal of TbMn$_2$O$_5$(100) with dimensions
of ($2\times1\times1$) mm$^3$ was cut and polished to achieve a
mirror-like surface, followed by a high-temperature annealing to
remove built-up strain during polishing. The modulation vector
$\vec{q}$ is the momentum transferred from the materials and lies
in the scattering plane defined by the $a$ and $c$ axes, i.e.,
$\vec{q}=(q_{a},0,q_{c})$. For photons of 639~eV, the intrinsic
$q$ resolution in terms of half width at half maxima (HWHM) is
estimated to be of 0.001~{\AA}$^{-1}$, including the effect of
photon penetration depth. The scattering results reveal that
TbMn$_2$O$_5$ exhibits an AF order below 42~K. Figure 1 shows the
energy dependence of scattering intensity with
$\vec{q}=(\frac{1}{2},0,\frac{1}{4})$ and HWHM, which reflects the
absorption of x-rays. To reduce self absorption, we used photons
with an energy just below the Mn $L_3$ edge in the following
discussion on the temperature and polarization dependence; for
example, the penetration depth ($\sim$~1800~\AA) of 639-eV photons
is much larger than the correlation length of AF order
($\sim$~800~{\AA} defined as the inverse of HWHM).

\begin{figure}
\includegraphics[width=8.5cm]{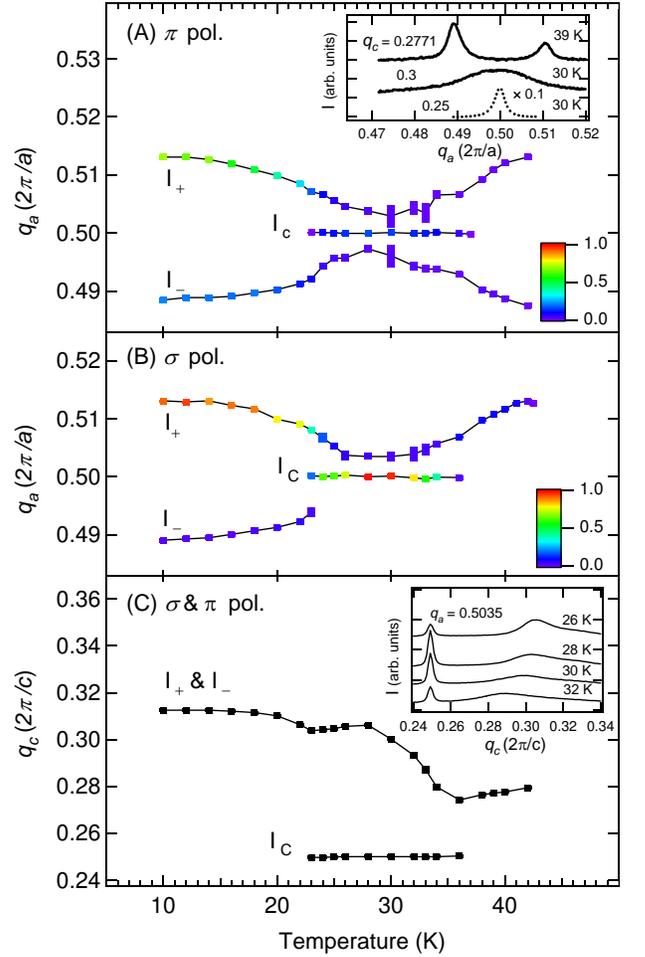}
\caption{(Color) Temperature-dependent $\vec{q}$ of AF order of
TbMn$_2$O$_5$: (A) and (B) $q_a$ component measured with $\pi$ and
$\sigma$ polarizations; (C) $q_c$ component. The vertical size of
rectangular symbols indicates the uncertainties of $q_a$ and
$q_c$. Intensities normalized to that of the commensurate AF order
at 30~K with $\sigma$ polarization are expressed by means of
color. The temperature dependence of $q_c$ is without showing
relative intensities. Intensities of incommensurate AF order with
$\vec{q}=(\frac{1}{2}{\pm}\delta_{a}, 0, \frac{1}{4}+\delta_{c})$
are denoted as $I_\pm$, and of the commensurate ordering as
$I_{\rm C}$. The insets of (A) and (C) are, respectively,
scattering intensities of $q_a$ and $q_c$ scans at selected
temperatures.}
\end{figure}

Temperature-dependent measurements indicate that the AF order of
TbMn$_2$O$_5$ occurs with modulation vectors $(\frac{1}{2} {\pm}
\delta_{a}, 0, \frac{1}{4}+\delta_{c})$, in which $\delta_a$ and
$\delta_c$ characterize the incommensurability. The temperature
dependence of $q_a$ is plotted in Fig. 2(A) and Fig. 2(B) for
$\pi$ and $\sigma$ polarizations, respectively. As the temperature
decreases, the incommensurate AF order of TbMn$_2$O$_5$ begins to
develop at 42~K, in agreement with neutron results
\cite{Kobayashi}. For the temperature between 37~K and 30~K, the
incommensurate scattering intensity decreases monotonically; the
$q_a$ of the incommensurate ordering moves toward 0.5. In
contrast, the commensurate ordering appears between 37~K and 24~K,
coexisting with the incommensurate ordering. The scattering
intensities plotted in the insets of Fig. 2(A) and Fig. 2(C)
demonstrate the coexistence of commensurate and incommensurate AF
orderings, similar to the coexistence of commensurate and
incommensurate AF phases observed in YMn$_2$O$_5$ \cite{Chapon_1}.
As shown in Fig. 3(A), the onset of spontaneous electric
polarization is accompanied by the incommensurate-commensurate AF
transition at 37~K, contrary to $R$MnO$_3$
\cite{Kimura,Kenzelmann, Arima}. On further lowering the
temperature, a commensurate-incommensurate transition occurs at 23
K; the commensurate ordering disappears.

The polarization dependence of x-ray scattering provides the
information about the direction of magnetic moments. Hannon
\emph{et al}. \cite{Hannon} have shown the magnetic moment along a
direction $\mathbf{\hat{Z}}$ probed in x-ray scattering is
proportional to
$(\mathbf{e}'^{\ast}\times\mathbf{e})\cdot\mathbf{\hat{Z}}$, where
$\mathbf{e}'$ is the electric field of the scattered light. For an
incident x-ray of $\sigma$ polarization, the scattered x-ray from
TbMn$_2$O$_5$ with $\mathbf{e}'{\parallel}b$ (denoted as $\sigma'$
polarization) makes no contribution to the magnetic scattering,
and the scattered x-ray with $\mathbf{e}' {\perp}b$ ($\pi'$
polarization) is predominantly sensitive to the magnetic moment
along the $a$ axis, i.e., $|S^a_q|$, because $\pi'^{\ast}\times
\sigma$ is $\sim 6^{\circ}$ away from the $a$ axis. Conversely, if
the incident x-ray has $\pi$ polarization, the scattered x-ray has
either $\pi'$ or $\sigma'$ polarization; $\pi'^{\ast}\times\pi$ is
parallel to the $b$ axis, whereas $\sigma'^{\ast}\times \pi$ is
predominantly along the $c$ direction. As neutron measurements
indicate that the magnetic moments are in the $ab$ plane
\cite{Kobayashi,Chapon}, the scattering with $\pi$ polarization is
predominantly sensitive to the magnetic moment along the $b$ axis,
i.e., $|S^b_q|$.

The knowledge of $|S^a_q|$ and $|S^b_q|$ enables one to
investigate how $\vec{P}$ is induced by magnetization based on the
Ginzburg-Landau approach
\cite{Kenzelmann,Lawes,Lawes05,Mostovoy06}. As hinted from the
broken inversion symmetry, there must be odd orders of $\vec{P}$
coupling to $\vec{S}_{\vec{q}}$. Clearly, the lowest order
coupling is that an internal field $\vec{E}_{in}$ couples to
$\vec{P}$, and the free energy $F$ can be written as
$F=P^2/2\chi_0 - \vec{E}_{in} \cdot \vec{P}$ with $\chi_0$ being
the electric susceptibility. The minimization of $F$ thus leads to
$\vec{P}= \chi_0 \vec{E}_{in}$. From symmetry point of view,
because $\vec{S}_{\vec{q}}$ changes sign under time reversal,
$\vec{E}_{in}$ must be quadratic in magnetization and contains at
least two components in $\vec{S}_{\vec{q}}$. Since both $\vec{q}$
and $-\vec{q}$ must be paired to make $\vec{P}$ uniform in real
space, $\vec{E}_{in}$ must contain both $\vec{S}_{\vec{q}}$ and
$\vec{S}_{-\vec{q}}$. Under the inversion operation  $\vec{r}_j
\rightarrow -\vec{r}_j$, $\vec{S}_{\vec{q}}$ becomes
$\vec{S}_{-\vec{q}}$. Since $\vec{P}$ changes sign, in order for
$F$ being invariant, $\vec{E}_{in}$ must change sign. Out of the
quantities that characterize the magnetic order and the underlying
lattice, there are two possible combinations for $\vec{E}_{in}$:
$\hat{u}\times(\vec{S}_{\vec{q}} \times {\vec{S}_{-\vec{q}}})$ or
$(\vec{S}_{\vec{q}} \cdot {\vec{S}_{-\vec{q}}})\hat{q}$, where
$\hat{u}$ is $\hat{a}$, $\hat{b}$, $\hat{c}$ or their combinations
\cite{Ref_Ein}, representing the most important anisotropic
direction in the spin-spin interaction. Since $(\vec{S}_{\vec{q}}
\cdot {\vec{S}_{-\vec{q}}})=|\vec{S}_{\vec{q}}|^2$ is finite for
any magnetic phase occurring below the Neel temperature $T_{\rm
N}$, an internal field of the form $(\vec{S}_{\vec{q}} \cdot
{\vec{S}_{-\vec{q}}})\hat{q}$ implies that transition temperature
for non-vanishing $\vec{P}$ is identical to $T_{\rm N}$, which
disagrees with experimental observation of TbMn$_2$O$_5$. On the
other hand, $\hat{u}\times(\vec{S}_{\vec{q}} \times
{\vec{S}_{-\vec{q}}})$ does discern different magnetic phases. In
particular, it vanishes for collinear or inversion-invariant
magnetic phases. Therefore, we conclude that $\vec{E}_{in} =
\sum_q i \gamma_{\vec{q}} \hat{u} \times ( \vec{S}_{\vec{q}}
\times {\vec{S}_{-\vec{q}}})$ is the only candidate that is
consistent with symmetry considerations for TbMn$_2$O$_5$. Here
$\gamma_{\vec{q}}$ is some unknown function to be determined by
microscopic models. Then the induced polarization $\vec{P}$ is
$\chi_0 \sum_q i \gamma_{\vec{q}} \hat{u} \times (
\vec{S}_{\vec{q}} \times {\vec{S}_{-\vec{q}}})$ \cite{Mostovoy06}.

\begin{figure}
\includegraphics[width=8.5cm]{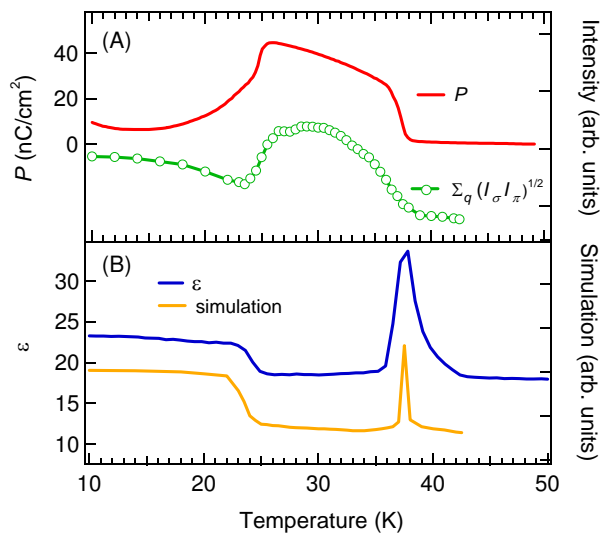}
\caption{(Color online) Temperature dependence of AF order and
ferroelectric properties of TbMn$_2$O$_5$: (A) Comparison of $P$
and intensities $\sum_{\vec{q}} ({\rm I}_{\sigma} {\rm
I}_{\pi})^{1/2}$; (B) Comparison of 1-kHz $\varepsilon$ along the
$b$ axis and simulation of $\chi_b$ with
$|S_q|^{2}=(I_\sigma+I_\pi)$. Here $\chi_{0}g_2(q_{IC})$~= 0.01,
$\chi_{0}g_2(q_{C})$~=$-0.01$, $g_1(q_{IC})$ = 10 and
$g_1(q_C)$=0. $\langle |\delta \vec{S}_q|^2 \rangle$ is simulated
with a Lorentzian distribution centring at 37.5 K with a HWHM of
0.5 K. Note that $I_\sigma$ and $I_\pi$ are the average
intensities for $\sigma$ and $\pi$ polarizations, respectively,
i.e., $I_{\sigma/\pi}\equiv(I_{+}+I_{-})/2$ for incommensurate
ordering and $I_{\sigma/\pi}\equiv I_{\rm C}$ for commensurate
ordering. $P$ and $\varepsilon$ are reproduced data of zero
magnetic field from Ref. \cite{Hur}.}
\end{figure}

For TbMn$_2$O$_5$, because $S_q^c=0$, by taking $\hat{u}=\hat{q}$,
$\vec{P}$ is along the $b$ axis, consistent with experimental
observations. The choice of $\hat{u}=\hat{q}$ is also consistent
with the work by Mostovoy \cite{Mostovoy06}, where the proposed
expression, after being averaged over space, also shows that
$\vec{P}$ is determined by $\vec{S}_{\vec{q}} \times
{\vec{S}_{-\vec{q}}}$. In addition, checking the magnitude of
$\vec{P}$ provides another justification. Since $P \propto
\sum_{\vec{q}} \gamma_{\vec{q}}|S^a_{\vec{q}}||S^b_{\vec{q}}|\sin
(\phi_a -\phi_b)$ \cite{pol} with $\phi_{a/b}$ being phases of
$S^{a/b}_{\vec{q}}$, the appearance of $\vec{P}$ requires
non-vanishing $\phi_a -\phi_b$, reflecting the requirement of
non-collinear spin structure for inducing polarization. Since the
change of $\vec{q}$ is small, $\gamma_{\vec{q}}$ is almost
temperature-independent. With further assumption of $\phi_a
-\phi_b$ being roughly independent of temperature, we have
$P\propto \sum_{\vec{q}} (I_{\pi}I_{\sigma})^\frac{1}{2}$. Figure
3(A) shows the comparison of $P$ with the sum of the intensities
$(I_{\pi}I_{\sigma})^\frac{1}{2}$ over commensurate and
incommensurate magnetic orderings. Clearly, they follow each other
closely, indicating the validity of the proposed $\vec{E}_{in}$
and the assumption on the temperature independence of $\phi_a
-\phi_b$. These observations also indicate that the coexisting
incommensurate and commensurate orderings break the inversion
symmetry. Furthermore, since $\vec{P}$ is an odd function in any
component of $\vec{S}_{\vec{q}}$, strong magnetic fields can
simply change the sign of just one component in
$\vec{S}_{\vec{q}}$ and result in the observed reversal of
$\vec{P}$ in direction \cite{Hur}.

For a complete understanding of the ferroelectricity, we further
discuss dielectric responses of TbMn$_2$O$_5$, as summarized in Fig.
3(B). Two anomalies are observed: $\varepsilon$ exhibits a sharp
maximum at the incommensurate-commensurate transition (37 K) and a
step-like structure at the commensurate-incommensurate transition
(23 K). As shown below, both anomalies reflect the
magneto-elasticity of the exchange energy. Specifically, because the
exchange energy $J$ is sensitive to positions of atoms
\cite{Sergienko}, applying $\vec{E}$ induces $\delta J$ that depends
on the change of polarization $\delta \vec{P}$ . Experimental data
indicate that only the component $E_b$ along the $b$ axis couples to
the magnetic order \cite{Hur} so that $\delta J$ only depends on
$\delta P_b$ and can be expressed as $\delta J=- g_1(q) E_b \delta
P_b + g_2(q) (\delta P_b)^2 /2$ with $g_1$ and $g_2$ being positive
numbers that characterize the corresponding couplings. Here the
first term contributes the potential energy of the electric dipole,
and when combined with $-\vec{E} \cdot \delta \vec{P}$, it corrects
$E_b$. The second term is the elastic energy that contributes a
correction to the original elastic energy $P^2 / 2 \chi_0$ and
changes $\chi_0$. Including these corrections, the electric
susceptibility becomes $\chi_b = \chi_0 (1+ \sum_q g_1(q) \langle
|\vec{S}_q|^2 \rangle) /(1+\chi_0 \sum_q g_2(q) \langle
|\vec{S}_q|^2 \rangle)$ \cite{DF}.

The expression of $\chi_b$ enables one to explain the behavior of
$\varepsilon$ phenomenologically. We first note that, for the
commensurate ordering at $\vec{q}_C$, our data indicates the
magnitude of $\vec{S}_{\vec{q}_C}$ saturates. It is then plausible
to assume that the exchange energy reaches maximum at commensurate
ordering, i.e., $ \partial \delta J / \partial \delta P_b =0$,
which enables one to eliminate $E_b$ and obtains $\delta J=-
g_2(q) (\delta P_b)^2 /2$. This is equivalent to set $g_1(q_C)=0$
and $g_2(q_C)<0$ from the beginning. Since $\langle |\vec{S}_q|^2
\rangle$ can be written as the sum of spin fluctuations, $\langle
|\delta \vec{S}_{\vec{q}}|^2 \rangle$, and square of moments,
$|\vec{S}_{\vec{q}}|^2$. The enhancement of spin fluctuations near
the incommensurate-commensurate transition explains the observed
sharp maximum \cite{Lawes} of $\varepsilon$ at 37 K. In contrast,
for the incommensurate ordering below 37 K, the magnitude of
$\vec{S}_{\vec{q}}$ is not saturated. In this case, since
$g_1(q_{IC})>0$ and $g_2(q_{IC})>0$, $\chi_b$ is thus more
sensitive to the numerator. Figure 3(B) shows the comparison of
the measured $\varepsilon$ along the $b$ axis and our
phenomenological simulation of $\chi_b$ with
$|S_{\vec{q}}|^2=(I_{\sigma}+I_{\pi})$. Note that $\chi_b$ differs
from $\varepsilon$ by a constant. The resemblance of these two
curves shows that the response to $\vec{E}$ indeed arises from
magneto-elasticity of the exchange energy.

We thank C.-H. Hsu, C. H. Chen, S. Ishihara, Y. Tokura,  G. Y.
Guo, and T. K. Lee for fruitful discussions and comments, and L.
L. Lee and H. W. Fu for their technical support. This work was
supported in part by the National Science Council of Taiwan. Work
at Rutgers was supported by NSF-MDR-0405682.


\begin{thebibliography}{99}
\bibitem{Ref_Ein}
Note that the possibility $\vec{S}_{\vec{q}} \times
{\vec{S}_{-\vec{q}}}$ is not considered because it does not couple
to the lattice; furthermore, its direction is also not consistent
with experiments.
\bibitem{pol} Because $\vec{S}_{\vec{q}}\times{\vec{S}_{-\vec{q}}}$
changes sign when $\vec{q}$ is changed to $-\vec{q}$ in the
summation, $\gamma_{\vec{q}}$ must be odd in $\vec{q}$ with the
lowest possible term being $\sin q_a$ or $\sin q_c$. Thus, for
$\gamma_{\vec{q}}=\sin q_a$, $P\propto\sum_q|S^a_q||S^b_q|\sin
q_a\sin(\phi_a -\phi_b)$, which is in agreement with microscopic
considerations \cite{Katsura,Sergienko}.
\bibitem{DF} The lowest-order terms for the change of $F$
can be written as $\delta F = [ \vec{P} / \chi_0 - i \sum_q \gamma_q
\hat{u} \times ( \vec{S}_q \times \vec{S}_{-q} ) ] \cdot \delta
\vec{P} + (\delta \vec{P})^2/2\chi_0 - \vec{E} \cdot \delta \vec{P}
+ \sum_q \delta J \langle \vec{S}_q \cdot \vec{S}_{-q} \rangle $.
Here the first term vanishes upon using the proposed relation
between $\vec{P}$ and $\vec{S}_q \times \vec{S}_{-q}$.
\end{thebibliography}
\end{document}